\documentstyle[12pt]{article}
\begin{document}

%..........Estos son Algunos Comandos Especiales que Pueden Usarse

\def\wh{\widehat}
\def\wt{\widetilde}
\def\D{\Cal{D}}
\def\ov{\overline}
\def\un{\underline}
\def\noi{\noindent}

%...................................................................

%\centerline {version 13 de junio de 1997}
%\centerline{\huge }
%\centerline{\huge  }
%\vskip 3cm
%\centerline{}
%\centerline
%\centerline{\it }
%\vskip .3cm
%\centerline{}
%\centerline
%\centerline{\it 

%\centerline{\it}
% Definition of title page: 
\title{Membrane Solitons  as Solitary Waves of Non-Linear Strings 
Dynamics}
\author{ {\Large {\it A. Restuccia$^1$ and R. Torrealba$^2$}}\\ \\ 
{\it $^1$ Depto. de Fisica, Universidad Simon Bolivar}\\ {\it Caracas 89000, 
Venezuela.}\\
{\it $^2$ Depto. de Matematicas , Decanato de Ciencias,}\\ {\it Universidad 
Centro Occidental 
``Lisandro Alvarado",} \\ { \it Barquisimeto, Venezuela.}
		% insert author(s) here
}
\date{}	% optional

\maketitle
\vskip .6cm
\begin{abstract}
Families of solutions to the field equations of the covariant BRST 
invariant effective action of the membrane theory are constructed.  
The equations are discussed in a double dimensional reduction, they 
lead to a nonlinear equation for a one dimensional extended object.  
One family of solutions of these equations are solitary waves with 
several properties of solitonic solutions in integrable systems, 
giving  evidence that in this double dimensional reduction the 
nonlinear equations are an integrable system.  The other family of 
solutions found, exploits the property that the non linear system 
under some assumptions is equivalent to a non linear Schr$\ddot 
{o}$dinger equation.
\end{abstract}

\newpage

\section{Introduction}
\noindent
 In recent years there has been a renewed interest in the theories
of Super-membranes and Dirichlet branes, specially due to the duality relations 
between many Super-D brane, Super-p branes and Superstring theories in 
several dimensions \cite{polchi}.  These duality relations has been 
stablished at 
the solitonic sectors
\cite{duff} of the low energy phenomenological actions.  Presumably there 
exists a fundamental 11 dimensional supersymmetric theory,
 from which all brane theories and its duality relations could be
obtained. In particular this theory must contain the solitonic sector of both 
Superstrings and its dual membranes.   Althought, a lot of progress in the
 understanding at the spectral 
\cite{bps}\cite{dewit}\cite{schwarz} level from the nonperturbative point of
 view has been reached,
there is still not satisfactory answer to fundamental questions as its geometrical
 principle, dynamics and quantum
consistency.
 One of the most striking difficulties of the bosonic p-brane theory  is
that the field equations are intrinsically non linear in all known gauges,
so it is not know how to solve the field equations at the classical level
neither how to perform its non perturbative quantization. This is 
the reason that almost all
 quantization aproach to p-branes or Super-p-brane are semiclassical
\cite{sezgin}.  We study in this paper some aspects of the non linear 
structure of the field equations of the covariant, BRST invariant, 
effective action of the Membrane Theory.
One way of having some insight into this nonlinear structure is to 
study sectors of the system which have integrability properties.  To 
do so we will consider a double dimensional reduction of the membrane 
theory.  We will then analyse the existence of solitary waves to this 
non linear sector.  The properties of the solitary waves give in 
general direct evidence of the existence or not of solitonic solutions to the 
non-linear system.  In fact, the properties of the solitary waves in systems 
like $KdV$, Sine Gordon, non-linear Schr$\ddot{o}$dinger equation and 
other members of their hierarchy give direct evidence of 
their solitonic behaviour.  These properties usually point out the 
way to perform the analysis of the non linear integrable equations by 
the inverse scattering method, which in most cases completely 
resolves the evolution problem of the non-linear system. 
In section 2 we briefly discuss the construction of the BRST invariant 
effective action of membrane theories, including the boundary 
conditions on ghost fields [7][8].  In section 3 we perform a double 
dimensional reduction leading to non linear field equations for a 
1-dim extended object.  In section 4 we find solitary wave solutions 
to it and discuss their properties.  In section 5, we consider the 
particular case of 4 dimensions and relate the problem to the static 
non linear  Schr$\ddot{o}$dinger equation.
 
\section{Off-shell BRST invariant effective action}

The off-shell BRST invariant effective action for the membrane can be
obtained
from the general modified BFV approach {\cite{yo}}. $x^\mu
,p_\mu$ denote the canonical conjugate bosonic variables, $ c^a, \mu_a $ are
the conjugate canonical ghost variables, $\lambda$ are the Lagrange
multipliers, $ {\un{c}}_a $ are auxiliary ghost fields and 
$\chi^a$ are the gauge 
fixing conditions.  The effective action is then

\begin{equation}
S_{eff}=\int{}dt\{p\dot{x}-\mu_a\dot{c}^a
-{\cal H}_0^{(brst)}-
\delta_{brst}(\lambda^a\mu_a)+\delta_{brst}({\un{c}}_a\chi^a)\}
\label{23}
\end{equation}

\noi where $\delta_{brst}(.)=[.,\Omega]$ and  $\Omega $ is the BRST
generator
\cite{brst}

\begin{equation}
        \Omega = c^a\phi_a + \mu _a {}^{(1)}U^a +  \mu _b\mu _a
{}^{(1)}U^{a,b}
\end{equation}
\begin{eqnarray}
        {}^{(1)}U^i & = & c^j\partial_j c^i -
\gamma\gamma^{ij}c^3\partial_jc^3,\nonumber \\
{}^{(1)}U^3 & = & c^3\partial_a c^a - \partial _a c^3 c^a,\nonumber \\
{}^{(2)}U^{ij} & = & \frac{3}{2}c^3\partial_jc^3 \partial_ic^3 \nonumber \\
        {}^{(2)}U^{3a} & = & 0.\nonumber
\end{eqnarray}

The functional integral obtained from ({\ref{23}}) reduces correctly to
the canonical
functional integral in terms of the physical modes with the correct
functional measure, and is independent of the gauge fixing
conditions \cite{brst}.

The action ({\ref{23}})  may be rewritten as

\begin{equation}
S_{eff} =\int{}d\sigma^3 p\dot{x}-\mu_a\dot{c}^a-
\lambda^a\phi_a{}^{(brst)}+
B^a\chi^a-\un{c}_a \delta_{brst}(\chi^a)-\theta^a\mu_a ,
\label{24}
\end{equation}

\noindent where we have introduced the non canonical auxiliary fields
$B^a$ and the
ghost auxiliary fields $\theta^a$, $\phi_a^{(brst)}=\{\mu_a,\Omega \}$
are the BRST
extension of the first class constraints

\begin{eqnarray}
        \phi _i &=& p_\mu x^\mu _i,  \\
        \phi _i &=& \frac{1}{2}(p^2 + \gamma ).
\end{eqnarray}

 From (\ref{24}) we obtain the Hamiltonian density as

\begin{eqnarray}
{\cal{H}}^{eff}=\lambda^a\phi_a{}^{(brst)}-B^a\chi_a+\un{c}_a
\delta_{brst}(\chi_a)+\theta^a\mu_a ,
\label{26}
\end{eqnarray}

\noindent and the effective BRST Lagrangean

\begin{equation}
{\cal{L}}^{eff}= p\dot{x}-\mu_a\dot{c}^a-{\cal H}^{eff}.
\label{27}
\end{equation}

Now we consider the covariant gauge fixing conditions

\begin{equation}
\chi^1:=\lambda^1 = 0,\ \ \ \chi^2:=\lambda^2=0,\ \ \ \chi^3:=\lambda^3-
1=0.
\label{28}
\end{equation}

\noindent after performing the functional integration on $B_a$ ,
$\theta^a$, and $\lambda^a$
in (\ref{24}) we obtain

\begin{equation}
{\cal{H}}^{eff}= \phi_3^{(brst)},
\label{29}
\end{equation}

\begin{equation}
S_{eff}=\int d\sigma^3 p\dot{x}-\mu_a\dot{c}^a-\phi_3^{(brst)},
\label{30}
\end{equation}

\begin{equation}
\phi_j^{(brst)}= \phi_j + (\partial_jc^a) \mu_a + \partial_i(c^i\mu_j) +
\partial_j(c^3\mu_3),
\label{31}
\end{equation}
\begin{equation}
\phi_3^{(brst)}= \frac{1}{2}(p^2+\gamma )+\gamma\gamma^{ij}
[\partial_jc^3 \mu_i] +
(\partial_ic^i)\mu_3+\frac{3}{2}\partial_ic^3\partial_jc^3\mu_j\mu_i+
\partial_i(c^i\mu_3).
\label{32}
\end{equation}

\noindent with periodic boundary conditions for the
closed membrane theory and the following boundary conditions for the
open membrane \cite{yo}:

\begin{equation}
        c^3|_{B}=0, \mbox{  and  } n_ic^i|_{B}=0,
        \label{bc}
\end{equation}

\noindent
(\ref{29}),(\ref{30}),(\ref{31}),(\ref{32}) agree with the results in 
\cite{fuji} which were obtained 
for the closed membrane only under several assumptions.

(\ref{bc}) are the generalization of the boundary conditions for the string 
\cite{kato}.  
Additionally the ghost fields must satisfy initial and final time
conditions,
 in order to obtain the BRST invariance of the action, they are:

\begin{equation}
c^3|_{t_i}=0,\mbox{  and  } c^3|_{t_f}=0,
\label{33}
\end{equation}

\noindent which are associated to the constraint quadratic in the momenta,
and are equivalent to the usual restriction on the gauge parameters
{\cite{teitel}}.

>From the effective Hamiltonian we get the field equations for the ghost
sector
\begin{eqnarray}
         && \dot{c}^i= \gamma\gamma^{ij}\partial_jc^3 +
         3\partial^ic^3\partial_jc^3\mu_j
        \label{cpuntoi}\nonumber \\
 && \dot{c}^3 = \partial_jc^j
        \label{cpunto3}\nonumber \\
        && \dot{\mu}_3 = \partial_i(\gamma\gamma^{ij}\mu_i) +
        3\partial_j(\partial_ic^3\mu_j\mu_i)
        \label{mupunto3}\nonumber \\
        &&\dot{\mu}_i = \partial_i\mu_3
        \label{mupuntoi}\nonumber
\end{eqnarray}

 \noindent We will consider the class of solutions satisfying the BRST 
 invariant condition $c_{3} = 0$ for all  $\sigma_1,\sigma_2,\tau$.  It is
 consistent with the previous boundary conditions. The field 
 equations for the ghosts
reduce then to

 \begin{equation}
        \dot{c}^i = 0, \mbox{ and } \partial_ic^i = 0
        \label{f.e}.
 \end{equation}

The field equations for the antighost are

\begin{eqnarray}
         &&\dot{\mu}_3 = \partial_j(\gamma\gamma^{ij}\mu_i)
        \label{mu3} \\
        &&\dot{\mu}_i = \partialÐi \mu_3
        \label{mui}
\end{eqnarray}

\noindent from which we obtain 

\begin{equation}
        \ddot{\mu}_3= \partial_j(\gamma\gamma^{ij}\partial_i\mu_3)
        \label{mudotdot}.
\end{equation}

\noindent
Finally the field equations for the bosonic coordinates become under 
our assumption

\begin{equation}
        \ddot{x}^\mu=\partial_j(\gamma\gamma^{ij}\partial_ix^\mu)
        \label{xdotdot}.
\end{equation}
\noindent
Once (\ref{xdotdot})  is solved (18) is a linear equation in $ \mu_{3}$.
Consequently it is enough to discuss the solutions of (\ref{xdotdot}) .  
We will 
construct in the next sections solitonic solutions to the covariant 
equation (\ref{xdotdot}) .
We will also consider solutions in the LCG. We notice that $X^{+}$, 
$X^{-}$ as in the LCG are solutions to (\ref{xdotdot}) .

\section{ Non Linear String Equation from D=11 Membrane Theory}
\noindent

One way of having some insight on the structure of a nonlinear dynamical 
system is
through the study of the solitary waves solutions of the system.  It usually 
happens that the existence of solitary waves leads to the existence of
 solitonic solutions of the system, which in turn allow through the 
 inverse scattering approach a complete
 understanding of at least an important sector of the full space of 
 solutions of the
 nonlinear system.

The structure of the solitary wave solutions we propose is closely related
to the decomposition of the 11 dimensional spacetime as the product 
$S^{1} \times M_{10}$ and in this sense we believe
that our solutions should be related to the IIA
Dirichlet Supermembranes. It is well known that the $D=11$ supermembrane 
with one 
coordinate of the target space compactified on $S_{1}$ is equivalent as 
a quantum field theory to the $D=10$ IIA Dirichlet supermembrane 
\cite{town}.
  We will consider the particular limit in which the radius of 
compactification of the membrane together with one of the world volume 
coordinates, also taken as an angular coordinate, are contracted to 
zero.  

We start analysing solutions with the following structure

\begin{equation}
        x^\mu = y(\sigma_2)\  f^\mu(\sigma_1 , \tau).
        \label{sigmaf}\\
\end{equation}

\noi
In this ansatz we separate the $\sigma_2$ dependence in order to
study the a dimensional reduction on one of the
target space time coordinates, which we will assume to have a $S^1 \times
M^{10}$ topology. It is well known that separation of variables do not
lead to a complete set of solutions of non linear equations as
(\ref{xdotdot}), but this ansatz is well suited for the double 
dimensional reduction we will shortly consider. 
Our argument is based on the remark that the field equations (\ref{xdotdot}) 
allow
$y(\sigma_2)$ to
be identified as a local coordinate over $S^1$.

Equation (\ref{xdotdot}) reduces after using (\ref{sigmaf}) to
\begin{eqnarray}
	 &y\ddot{f}^\mu - y(y')^2[{f^\mu}_1(f^2)]_{,1} - 
[y'f^\mu y^2{f_1}^2]_{,2} 
	  & \nonumber \\
	 &  & 
	\label{3.2} \\
	 & + [y(y')^2 (f.f_1) f^\mu]_{,1} + [y^2 y'(f.f_1) {f^\mu}_1]_{,2} & =0 \nonumber
\end{eqnarray}

\noi which after some manipulations yields
\begin{eqnarray}
\ddot{f}^\mu = &(y')^2\left([{f^\mu}_1f^2]_{,1} + 2 f^\mu f_{1}^{2} -[(f.f_1)f^
\mu]_{,1}-2(f.f_{1})f^{\mu}_{1}\right)  &  \nonumber\\
	 &  & 
	\label{3.3} \\
	 & + yy''[f^{\mu}f_{1}^{2} - (f.f_1) f^{\mu}_{1}] & \nonumber
\end{eqnarray}

In standard Kaluza Klein compactifications it is assumed a Fourier expansion
for the local coordinate on a compact manifold, but in this case
$y(\sigma_2)=\kappa^{(n)} e^{in/R \sigma_2}$
leads to different orders in the $(y')^2$ and $(yy'')$ terms, it may also
be meaningless to take linear combinations to study nonlinear equations, so it
is better to consider each of the n-th modes by taking a polinomial 
dependence in
$\sigma_2$

\begin{equation}
        y(\sigma_2)= \kappa_{(n)} \sigma_2^n ,\nonumber
\end{equation}

\noi
then  $(y_{(n)}')^2 = n^2 \kappa_{(n)}^2 \sigma_2^{2n-2}$  and
 $y_{n}y''_{n} = n(n-1) \kappa_{(n)}^2 \sigma_2^{2n-2}$ that have the same
order.

So, to factorize the $\sigma_2$ dependence in (\ref{3.3}) we must have
$n^2=n(n-1)$ this implies that $n=0$ and $n=1$.
Then it is natural to take  $y(\sigma_2)$ of the form

\begin{equation}
        y(\sigma_2)=\kappa\sigma_2 + c
        \label{3.4}
\end{equation}
where $\kappa$ and $c$ are constants.

Using (\ref{3.4}) equation, (\ref{3.3}) reduces to the nonlinear equation

\begin{equation}
        \ddot{f}^{\mu}=\kappa^2(f^{\mu}_{11}f^2 + f^\mu{f_1}^2
-(f_{11}.f)f^{\mu}-(f_1.f){f^\mu}_1)
        \label{efe}
\end{equation}

 Let us take $\theta$ on the interval $ [0,2\pi]$ as the angular
coordinate on $S^1$, and let $\sigma_2$ be
a world sheet coodinate wrapped around a circle of radius $r$, so 
that $\sigma_{2}=r\theta$. 
We will consider $x^{11}$ to be wrapped wround a circle of radius $ R = 
O(r) r \kappa $ with $O(r) \rightarrow 0$ when $r \rightarrow 0$, and 
c as the initial lenght on the circle: $c=2\pi N R$.  Now we perform a 
double dimensional reduction on the target and the worldsheet spaces, 
we take $ N \rightarrow \infty$ keeping c constant.  Then $ R \rightarrow 
0$ which implies $ r \rightarrow 0$.  We thus obtain
\begin{equation}
        x^{11} \rightarrow 0,\nonumber
\end{equation}

\begin{equation}
x^{u} \rightarrow c f^u(\sigma_1,\tau),\mbox{    where u=1,...,10}
\label{x^u}
\end{equation}

So, now  $x^u$ are the coordinates of a 10 dimensional string
that obeys the nonlinear equation (\ref{efe}).

The double dimensional reduction we have performed is different
from the one in \cite{town}. However the
 non linear string equation (25) is clearly related to the field 
 equations of the Dirichlet branes
 in ten dimension.  The membrane theory, in this limit, may then 
 be interpreted as the interaction of 10-dim strings.

 In the next section we will study the solitary wave solutions of the
nonlinear
 equation (\ref{efe}). We will perform the analysis in $D$ dimension and
discuss
 afterward the particular case $D=4$.

\section{ Solitary wave solutions}
\noindent
Now we try a solitary wave solution

\begin{equation}
        f^\mu = f^{\mu}(\sigma_1 - v\tau)\, ,
        \label{sol}
\end{equation}

\noi then (\ref{efe}) may be rewritten as

\begin{equation}
        (v^2-f^2){f^\mu}_{11} - f^\mu({f_1}^2 -f_{11}.f) +
(f_1.f){f^\mu}_{1}=0\, .
        \label{feq}
\end{equation}
where we have taken $\kappa=1$
\noi.

 After multiplying by $f^{\mu}_{1}$, and some rearrangements
we obtain 

\begin{equation}
        \frac{1}{2}[(v^2-f^2)\ {f_1}^2]_1 + \frac{1}{8}[\{(f^2)_1\}^2]_1 =
0\, .
        \label{const}
\end{equation}

\noi From which 

\begin{equation}
        (v^2-f^2)f_1^2 + \frac{1}{4}[(f^2)_1]^2 = C
        \label{C}
\end{equation}

\noi so we can solve for $f{^2}_{1}$ and introduce back it into (29), and after
 some calculations we get 

\begin{equation}
        (v^2-f^2)
        [-(f_1)^2 + \frac{1}{2}(f^2)_{11}]
        -2f^2
        \frac{C-\frac{1}{4}[(f^2)_1]^2}{v^2-f^2} +
        \frac{1}{2}f^2(f^2)_{11}
        +[\frac{1}{2}(f^2)_1]^2=0\, .
        \label{f^2}
\end{equation}

 \noi Defining  $u= v^2 - f^2$ , it is straighforward to obtain

\begin{equation}
        -2C + u \frac{C}{v^2} + \frac{1}{2}{u_1}^2 - \frac{1}{2} u
u_{11} = 0.
        \label{u}
\end{equation}

When $ C=0$ (\ref{u}) may be integrated giving

\begin{equation}
         B e^{k(\sigma - v\tau)} = f^2 - v^2\, .
        \label{solu}
\end{equation}

\noi Introducing the latest equation into (\ref{feq}) we obtain

\begin{equation}
        -{f^\mu}_{11} + \frac{1}{2}k{f^\mu}_{1}=0\, ,
        \label{f11}
\end{equation}

\noi hence the general solitary wave solution with $C=0$ is

\begin{equation}
        f^\mu = B^\mu e^{\frac{1}{2}k(\sigma-v\tau)} + A^\mu
        \label{f}
\end{equation}

\noi where $ (B^\mu )^2 = B $, $ (A^\mu )^2 = v^2 $, and $ B^\mu .
A^\mu  = 0$.
When $C$ is not null, we can obtain the solution to (33) by considering

\begin {equation}
        u_{c} = B e^{k( \sigma-vt)} + 2v^2
        \label{uc}
\end{equation}

\noi where $k = {\frac{\sqrt {C}}{v^2}}$. Using $u_c = v^2 - f^2$ we obtain
$f^2$ and
its derivatives, and introducing these into (32) we get a linear ordinary 
differential 
equation for the coordinates $f^{\mu}(\sigma)$

\begin{equation}
u_{c}{{f^\mu}}_{11} - \frac{1}{2}(u_{c})_{1} {{f^\mu}}_{1} -
k^{2}v^{2}{{f^\mu}}= 0, 
\label {lineal}\\
\end{equation}
 where $\mu$ = 1,...,9 and $u_{c}$ is given by (\ref{uc}).
\noi This implies that one may find all the solutions of the type
(\ref{sigmaf}) which
 are solitary waves by solving the linear differential equation 
 (\ref{lineal}), together with the consistency condition 
 $u_{c}=v^{2}- f^{2}$.
 
Note that the amplitude of this solitary wave solution is velocity
dependent, this is a strong evidence that this solution must be a
soliton for the classical bosonic membrane.

 It is also interesting to see
that all the solutions with $c=0$ and $c\neq 0$ are non degenerate, 
in the sense that its area at
constant time folding of the space-time is not zero $(\gamma \neq 
0)$. 
The solutions (\ref{f})(\ref{lineal}) have several properties in common
 with the solitonic 
solutions of integrable systems like KdV, Sine-Gordon, Non Linear 
Schr$\ddot{o}$dinger Equation. The amplitude, A and B in (\ref{f}), 
depends on the velocity of propagation v.  The wave number is v 
dependent($C\neq 0$).  There are solitary waves solutions for any value of v 
while the non linear equation (\ref{xdotdot}) is independent of v. 
There is a balance between the dispersion relation and the 
nonlinearity of the system  allowing solitary wave solutions.

 For a 
particular value of v (\ref{f}) represents a solution of the 
Nambu Goto string 
fields equations.  This string solution represents a ``Soliton" of the 
membrane equations.

We can also discuss following the same lines solutions to the field
 equations of the  
Fujikawa [11] effective action.  In that case there are different 
boundary conditions and the following gauge conditions have to be 
imposed:

\begin{eqnarray}
        g^{oo} + \gamma =0 \nonumber \\
        g_{oi} = x_{o}.x_{i}=0
        \label{goi}
\end{eqnarray}

then in order to have a solution, (\ref{f}) has to satisfy further 
restrictions which lead to 

\begin{eqnarray}
         && -\frac{1}{4}k^2 v (\kappa\sigma_{2}-c)^2 B e^{k(\sigma_1 - v
         \tau)}=0 \nonumber\\
        && -\frac{1}{2}k v (\kappa\sigma_{2}-c) B e^{k(\sigma_1 - v
         \tau)}=0.\nonumber\
        \label{B}
\end{eqnarray}

\noindent Since $k$ and $v$ are not zero, then $B=0$, and 

\begin{eqnarray}
         \gamma & = &det x_{i}x_{j}
        \nonumber \\
        \gamma & = &
\frac{1}{4}Bk^2v^2(\kappa\sigma_2-c)e^{k(\sigma_{1}-v\tau)}=0
        \label{gamma}
\end{eqnarray}

This implies then, that for the Fujikawa effective theory we obtain solutions 
describing a degenerated membrane
 with no area. These degenerate non
linear string solitonic solutions are then solutions of the bosonic membrane
theory even without  performing any dimensional reduction.
We will obtain now solitary wave solutions of equations (\ref{xdotdot})
satisfying also Fujikawa conditions with non zero area.
 We do so by working in the light cone gauge (LCG). As explained in \cite{hope}
we take

\begin{equation}
          x^{+} = c^+\tau
        \label{lcg}
\end{equation}

\noindent then the gauge fixing condition (\ref {goi}) allows us to solve for
the
minus coordinate

\begin{equation}
        x^{-}_{,i} = \frac{1}{c^+} \dot{ \vec{x}} \cdot \vec{x_{i}}
        \label{37}
\end{equation}

 \noindent in term of the LCG tranverse sector.  The effective action and the
field
equations are the same as before (see (10)-(12) and (19)) upon replacement
of $x^\mu $ by  $\vec{x}$
and the area expanded by the tranverse solitary  wave solution
is not zero because now (\ref{goi}) implies (\ref{37}) and then
neither $ B$
not   $ \gamma $ are null.
If we perform a double dimensional reduction as explained in section(3),
we end up with a non linear string solution $x^u = c f^u(\sigma_1,\tau)$
where $u = 1,...,8$  and with the usual LCG prescription for $x^{+}$ and
$x^{-}$.  

\section{New membrane solutions in 4 dimensions}

\noindent  Very interesting solutions for the bosonic membrane arise
in four dimension. We again choose the conformal (\ref {28}) and LCG
(\ref {lcg}) fixing. Denoting

\begin{equation}
        x^\mu = ( x^+, x^-,x^1,x^2)
        \label{xmu}
\end{equation}

\noindent we will use complex variable to represent the LCG tranverse
sector

\begin{equation}
        Z = x^{1} + i x^{2}
\end{equation}

\noindent and look for solutions of the  type

\begin{equation}
        Z = y(\sigma _{2}) F(\sigma _{1}, \tau).
        \label{yf}
\end{equation}

Equation (\ref{efe}) may be written as

\begin{eqnarray}
\ddot {F}&=& (\partial_{2}y )^2\{(F_{1}|F|^2)_{1} + 2F|F_1|^2 -
\frac{1}{2}[(F_{1}\bar{F} +\bar{F}_{1}F)F]_{1}
 - (F_{1}\bar{F}+\bar{F}_{1}F)F_{1}\}
\nonumber\\
&&-\frac{1}{2}\ y(\partial_{22}y)\ [F^2_{1}\bar{F}-
|F_{1}|^2F],
\label{Fea}
\end{eqnarray}

\noindent This equation allows as before, the solution if
$y=k\sigma_{2}$
and  from (\ref{Fea})  we get

\begin{equation}
 k^2 = \frac{\ddot F}{{(F_{1} |F|^2)_{1} +2F |F_1|^2 -
\frac{1}{2} [(F_{1}\bar{F} + \bar{F}_{1} F) F]_{1} - (F_{1}\bar{F} +
\bar{F}_{1} F) F_{1}}}.
        \label{Feisima}
\end{equation}

\noindent lets take $k=1$ and try  $F = X(\sigma_{1}) T(\tau)$ both
complex functions, then (\ref{Feisima}) completely separates as:

\begin{equation}
        \frac{\ddot T}{T |T|^2} = \frac{1}{2} X_{11} \bar X
        + \frac{1}{2}|X|^2 - \frac{1}{2} X{\bar{X}}_{11}
        -\frac{1}{2}(X_{1})^2\frac{\bar{X}}{X} = \frac{A}{2}
        \end{equation}

The $\sigma_1$ depending equation is

\begin{equation}
        AX - X_{11} |X|^2 - |X_1|^2 X +  X^2 {\bar{X}}_{11}
        +(X_{1})^2 \bar{X} = 0
        \end{equation}

\noindent
and may be solved introducing polar coordinates
$X=r(\sigma_1)e^{i\theta(\sigma_1)}$ that reduce the equation to:

\begin{equation}
\frac{A}{2}=ir(r\theta{}')'+(r\theta{}')^2
\end{equation}

As $A$ is a constant,
this implies that $r\theta {}' = 0$
then $Im[A] = 0$ so $A$ must be a real nonegative number and $\theta$
can be directly integrated

\begin{equation}
\theta=\sqrt{\frac{A}{2}}\int\frac{d\sigma_1}{r(\sigma_1)}\\
\end{equation}

\noindent where $r(\sigma_1)$ is an arbitrary real function.

The time depending equation is a non linear static Schrodinger
equation \cite{scott} upon interchange of $\tau$ and $\sigma_1$.

\begin{equation}
\ddot T = \frac{A}{2} T |T|^2 ,
        \label{T}
\end{equation}

\noi
this equation can be solved, as usual, trying with a function that
is product of a phase function by an envelope function

\begin{equation}
T=h(t)e^{i\phi(t)}
\end{equation}

\noi
splilting (\ref {T}) into real and imaginary parts, we get

\begin{eqnarray}
&&\ddot{h}-h\dot{\phi}^2=\frac{A}{2}h^3
\label{h}\\
&&h\ddot{\phi}+2\dot{h}\dot{\phi} = 0
\end{eqnarray}

\noindent Equation (\ref{h}) may be integrated to give 

\begin{equation}
\phi{}=\int{}dt\frac{C}{h^2(t)}
\label{phi}
\end{equation}

\noindent if $C=0$, then $T$ is real and the ( \ref{T})  reduces to
 the  well known case of a particle submited to a cubic force. This
is easily solve by  a quadrature that leads to first class elliptic
Jacobi functions \cite{libro}. Also when $C$ is not null the (\ref{h})
 (real part of (\ref{T})) can be solved by more complicated
elliptic Jacobi functions.

Using (\ref{phi}) into (\ref{h}) we obtain a quadrature from which

\begin{equation}
        \dot{h}^2 = \frac{1}{4}(B h^2 + A h^6 - C^2 h^{-2})
\end{equation}

\noi making the variable change $ u = h^2$ the latest equation yields 

\begin{equation}
t=\int\frac{du}{\sqrt{A u^3+Bu-C^2}}
\label{te}
\end{equation}

\noi
that is one Weirstrass elliptic function, where $A$, $B$, and $C$ are
constants and $u(t)$ could be obtained by
inverting $t(u)$ as an Jacobi elliptic function \cite{libro}.

\section{Conclusions}
\noindent Starting for the BRST invariant effective action for the membrane 
theory we found new solutions to the non-linear field equation.  The analysis 
was performed in a double dimensional reduction leading to a nonlinear field 
equation for a 1-dim extended object.  In this reduction the membrane 
represents then an interacting theory of 10-dim strings, in agreement with 
\cite{russo}.

We constructed the general solitary wave satisfying that non-linear 
differential system.  The solitary waves present similar properties to the 
solitonic  solutions on integrable systems like $KdV$, Sine Gordon, 
and other members of their hierarchy.  It is natural to conjeture 
that this solutions are true solitons of the non linear field 
equations giving then strong evidence that the system is integrable 
in the sense that can be completely resolved by using the inverse 
scattering method.  Pointing in that direction, a particular 
construction was presented in 4-dim relating the non-linear system to 
the non-linear  Schr$\ddot{o}$dinger equation.

\end{document}